\documentclass[prl,twocolumn,superscriptaddress,floatfix]{revtex4-1}

\usepackage[sort&compress]{natbib}
\usepackage{graphicx, times}
\usepackage{color}

\usepackage{subeqnarray}
\begin{document}
\title{Ultrafast Quantum Gates in Circuit QED}
\author{G. Romero}
\affiliation{Departamento de Qu\'{\i}mica F\'{\i}sica, Universidad del Pa\'{\i}s Vasco-Euskal Herriko Unibertsitatea, Apartado 644, 48080 Bilbao, Spain}
\author{D. Ballester}
\affiliation{Departamento de Qu\'{\i}mica F\'{\i}sica, Universidad del Pa\'{\i}s Vasco-Euskal Herriko Unibertsitatea, Apartado 644, 48080 Bilbao, Spain}
\author{Y. M. Wang}
\affiliation{Centre for Quantum Technologies, National University of Singapore, 3 Science Drive 2, Singapore 117543}
\author{V. Scarani}
\affiliation{Centre for Quantum Technologies, National University of Singapore, 3 Science Drive 2, Singapore 117543}
\affiliation{Department of Physics, National University of Singapore, 2 Science Drive 3, Singapore 117542}
\author{E. Solano}
\affiliation{Departamento de Qu\'{\i}mica F\'{\i}sica, Universidad del Pa\'{\i}s Vasco-Euskal Herriko Unibertsitatea, Apartado 644, 48080 Bilbao, Spain}
\affiliation{IKERBASQUE, Basque Foundation for Science, Alameda Urquijo 36, 48011 Bilbao, Spain}

\begin{abstract}
We present a method to implement fast two-qubit gates valid for the ultrastrong coupling~(USC) and deep strong coupling~(DSC) regimes of light-matter interaction, considering state-of-the-art circuit quantum electrodynamics (QED) technology. Our proposal includes a suitable qubit architecture and is based on a four-step sequential displacement of the intracavity field, operating at a time proportional to the inverse of the resonator frequency. Through ab initio calculations, we show that these quantum gates can be performed at subnanosecond time scales, while keeping a fidelity above $99\%$.
\end{abstract}

\date{\today}

\maketitle

{\it Introduction}.---With the advent of quantum information science~\cite{NC}, there have been enormous efforts in the design of devices with high level of quantum control and coherence~\cite{QC}. Circuit QED~\cite{Blais04,Wallraff04,Chiorescu04} has become a leading technology for solid-state based quantum computation and its performance is approaching that of trapped ions~\cite{IonsReview} and all-optical implementations~\cite{Optical}. Considerable progress has been made in recent circuit QED experiments involving ultrastrong coupling \cite{Niemczyk10,Pol10}, two-qubit gate and algorithms \cite{Plantenberg07,DiCarlo09,Yamamoto10,Bialczak10,Geller10,Haack10}, three-qubit gate and entanglement \cite{3qbit, MM, Wallraff_te}. Most of the proposed gates are based on slow dispersive interactions or faster resonant gates, and would require operation times of about tens of nanoseconds.

To speed up gate operations, the latest circuit-QED technology offers the USC regime of light-matter interactions~\cite{Ciuti05,Bourassa09,Niemczyk10,Pol10,Borja10}, where the coupling strength $g$ is comparable to the resonator frequency~$\omega_r$ ($0.1\lesssim \! g/ \omega_r\lesssim \! 1$). This should open the possibility to achieve fast gates operating at subnanosecond time scales~\cite{JJ1,JJ2} . In this sense, the design of these novel gates becomes a challenge as the rotating-wave approximation (RWA) breaks down and the complexity of the quantum Rabi Hamiltonian emerges~\cite{Braak11,Solano11}. Preliminary efforts have been done in this direction involving different configurations of superconducting circuits~\cite{Wang04,Wang09,Wang10}. Likewise, in a recent contribution, it has been discussed the possibility of performing protected quantum computing~\cite{Ciuti11}.

\begin{figure}[b]
\includegraphics[width=0.4\textwidth]{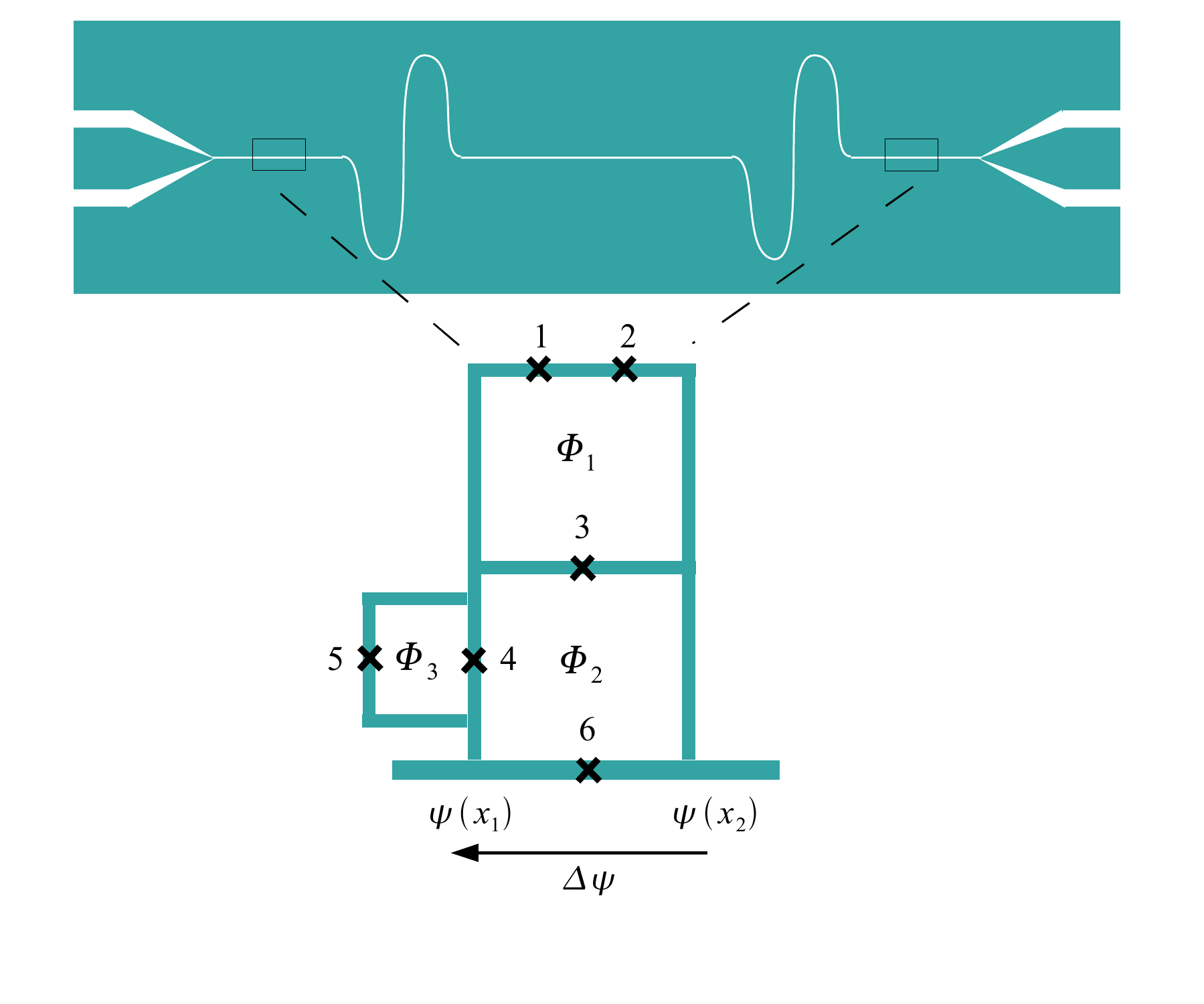}
\caption{\label{fig1} (Color online) Circuit-QED configuration of a six-Josephson-junctions array coupled galvanically to a resonator (bottom line). The flux qubit is defined by three Josephson junctions in the upper loop threaded by an external flux $\Phi_1$. Two additional loops allow a tunable and switchable qubit-resonator coupling.}
\end{figure}

In this Letter, we propose a realistic scheme to realize fast two-qubit controlled phase (CPHASE) gates between two newly designed flux qubits~\cite{Mazo99}, coupled galvanically to a single-mode transmission line resonator (Fig.~\ref{fig1}). Our proposal includes: (i) a CPHASE gate protocol operating at times proportional to the inverse of the resonator frequency; (ii) the design of the qubit-resonator system, allowing for high controllability on both the qubit transition frequency and the qubit-resonator coupling, in USC~\cite{Niemczyk10,Pol10} and potentially the DSC regime~\cite{Casanova2010} of light-matter interaction. Through ab initio numerical analysis, we discuss the main features of this scheme in detail and show that the fidelity could reach $99.6\%$. This is an important step in the reduction of resources requirement for fault-tolerant quantum computation~\cite{Fault}.

{\it Design of a versatile flux qubit}.---The junction array is schematically depicted in Fig.~\ref{fig1}. It consists of a six-Josephson-junctions configuration, each one denoted by a cross, coupled galvanically~\cite{Niemczyk10,Pol10,Abdumalikov08} to a coplanar waveguide resonator. This is characterized by a superconducting phase difference $\Delta\psi$ and the electromagnetic field supported. The upper loop stands for a three-junction flux qubit~\cite{Mazo99}, while the additional loops will allow a tunable quit-resonator coupling strength. We analyze this qubit design step-by-step. First, we describe the potential energy coming from the inductive terms, which is the dominant contribution. Second, we introduce the transmission line resonator inserted with a Josephson junction in the central line and introduce the superconducting phase slip $\Delta\psi$ across the junction shared with the resonator. Third, we add the capacitive terms which appear in the junctions and obtain the full Hamiltonian of the system using a standard procedure. Finally, we identify two levels in the degrees of freedom of the junction architecture, which will define our qubit, and obtain the effective Hamiltonian to describe the fast two-qubit gate.

The potential energy due to the inductive terms is obtained by adding up the corresponding Josephson potentials ${\cal E}(\varphi_k)=-E_{Jk}\cos(\varphi_k)$, where $E_{Jk}$ and $\varphi_k$ represent the Josephson energy and the superconducting phase across the $k$-th junction, respectively. We assume $E_{J1}=E_{J2}\equiv E_J$, $E_{J3}=\alpha E_{J}$ and $E_{J4}=E_{J5}=\alpha_4 E_J$. In addition, around each closed loop the total flux has to be a multiple of the flux quantum $\Phi_0=h/2e$, or expressed in terms of superconducting phases, $\sum_k\varphi_k=2\pi f_j+2\pi n$, where we defined the frustration parameter $f_j=\Phi_j/\Phi_0$~($j=1,2,3$). Using this quantization rule, the total potential energy $U$, containing both the qubit energy and the qubit-resonator interaction, reads
\begin{eqnarray}
\frac{U}{E_{J}}&=&-[\cos\varphi_1 + \cos\varphi_2 + \alpha \cos(\varphi_2 - \varphi_1 + 2\pi f_1)\nonumber\\
&+&2\alpha_4(f_3)\cos(\varphi_2-\varphi_1+2\pi\tilde{f} + \Delta\psi)]\,,
\label{Energy_potential}
\end{eqnarray}
where $\alpha_4(f_3)\equiv\alpha_4\cos(\pi f_3)$, $\tilde{f}=f_1 + f_2 + f_3/2$, and $\Delta \psi$ stands for the phase slip shared by the resonator and $f_2$-loop. Note that the junction at the central line introduces a boundary condition that modifies the mode structure of the resonator, but without altering the inductive potential~(\ref{Energy_potential}). The parameters $\alpha$, $\alpha_4$ and $f_j$ can be optimized to find a suitable working point.

We introduce now the Hamiltonian of the interrupted inhomogeneous transmission line resonator. After mapping the nonuniform resonator into a sum of harmonic oscillators, the standard Hamiltonian described by the flux amplitude $\psi_n$ and charge $q_n$ as its conjugate momentum is ~\cite{Bourassa09}
\begin{eqnarray}
{\cal H}_r=\sum_n \left( \frac{q_n^2}{2 \tilde{C}_r}  + \frac{\tilde{C}_r}{2} \omega_n^2 \psi_n^2 \right),
\end{eqnarray}
where $\tilde{C}_r = C_r + C_{J6}$ is the modified resonator capacitance due to the presence of the sixth Josephson junction at the central line, see Fig.~\ref{fig1}. The quantization procedure yields
\begin{eqnarray}
{\cal H}_r=\sum_n \hbar \omega_n \left(a_n^+ a_n +1/2 \right),
\end{eqnarray}
with $\omega_n=k_n/\sqrt{L_0 C_0}$ ($C_0, \,L_0$ the capacitance and inductance per unit length, respectively), and
$k_n=(2L_0/L_J) (1-\omega_n^2/\omega_p^2) \cot(k_n l)$, where $\omega_p=1/\sqrt{L_J C_J}$ is the plasma frequency of the junction and $2\, l$ is the length of the central line.

For simplicity in the presentation, assume that the junctions only interact with the first eigenmode of the resonator. The modified superconducting phase slip $\Delta\psi$ is related to the single-mode resonator variables as
\begin{equation}
\Delta \psi =\Delta{\psi_1} (a + a^{\dag})\,,
\end{equation}
where $\Delta{\psi_1}=(\delta_1/\varphi_0)(\hbar/2\omega_r\tilde{C}_r)^{1/2}$. Here, $\varphi_0=\Phi_0/2\pi$ is the reduced flux quantum, $\omega_r$ is the frequency of the first eigenmode, $\delta_1=u_1(x_2)-u_1(x_1)$ corresponds to the difference between the first-order spatial eigenmode, evaluated at points shared by the resonator and the $f_2$-loop.

The total Hamiltonian is obtained by including the kinetic terms of the qubit and the Hamiltonian of the resonator
\begin{eqnarray}
{\cal H}&=& 4 A E_c (n_1^2 + n_2^2) + 8 B E_c n_1  n_2 \,+\, {\cal H}_r \nonumber\\
&+& 2 e \frac{C}{C_r} q_n (n_1-n_2) \,+\,{U}(\varphi_1,\varphi_2),
\label{fullham}
\end{eqnarray}
where $E_c=e^2/{2 C_J}$ is the charging energy, $A,B,C$ are functions of system parameters~\cite{ABC}, such as junctions size and phase slip magnitude. The degrees of freedom of the junction architecture are $(\varphi_1,\varphi_2)$ and their conjugate momenta, which are the numbers $(n_1,n_2)$ of Cooper pairs in the boxes.

In order to define a qubit within the junction architecture, we diagonalize the term of the Hamiltonian containing only the junctions. The two lowest energy eigenstates are labeled as the eigenstates of $\sigma_z$, and the two-dimensional subspace spanned by them describes the qubit. Furthermore, since $\Delta\psi_1\ll1$ in general, we expand the potential (\ref{Energy_potential}) up to the second order in $\Delta\psi$. This gives rise to the first-order and second-order qubit-resonator inductively coupling. Finally, after projecting the interaction terms also into the qubit basis, the Hamiltonian reads
\begin{eqnarray}
{\cal H} &=& \frac{\hbar\omega_q}{2}\,\sigma_z + \hbar \omega_r a^{\dag}a + {\cal H}_{\rm int}\label{long_Hamil}
\end{eqnarray}
with the effective interaction Hamiltonian
\begin{eqnarray}
{\cal H}_{\rm int}=2E_{J}\alpha_4(f_3)\sum_{m={1,2}}(\Delta\psi)^m \sum_{\mu={x,y,z}}c^m_{\mu}(\alpha,\alpha_4,f_1,f_2)\sigma_\mu,\nonumber \\
\label{eff_Hamil}
\end{eqnarray}
$c^m_{\mu}(\alpha,\alpha_4,f_1,f_2)$ being the controllable magnitudes of the longitudinal and transverse coupling strengths for $m$-th order interaction. Here we have ignored the capacitive coupling, as it is orders of magnitude smaller than the inductively coupling.

{\it Numerical analysis}.---We provide ab initio numerical examples to show the functionality of our setup.
Firstly, we study the properties of the inhomogeneous transmission line resonator obtained by inserting the sixth Josephson junction on the central line. The eigenfrequencies $\omega_n$ and eigenmodes $u_n(x)$ can be found numerically and calculated with independence of the qubit Hamiltonian~\cite{Bourassa09}. In fact, considering a resonator of impedance $Z\sim50~\Omega$, capacitance $C_r\sim850~{\rm fF}$, and a Josephson capacitance $C_{J6}\sim10~{\rm fF}$, we estimate the first mode of the resonator with frequency $\omega_r/2\pi\sim 7~{\rm GHz}$, which leads to a phase slip magnitude $\Delta\psi_1=0.1218$. These values determine the qubit-resonator coupling strength. Secondly, we perform the numerical study of the total Hamiltonian~(\ref{fullham}), which shows that when external fluxes satisfy $f_2+f_3/2=0.5$, both $c^1_y$ and the second-order coupling are negligible. This reduces the interaction Hamiltonian to
\begin{equation}
{\cal H}_{\rm int} = \hbar g (a + a^{\dag})(c_z\sigma_z+c_x\sigma_x),
\label{long_Hami2}
\end{equation}
with the effective coupling strength $g=2E_J\,\alpha_4(f_3)\,\Delta\psi_1/\hbar$ and $c_{z,x}\equiv c^{1}_{x,z}$.

\begin{figure}
\includegraphics[width=0.23\textwidth,height=0.2\textwidth]{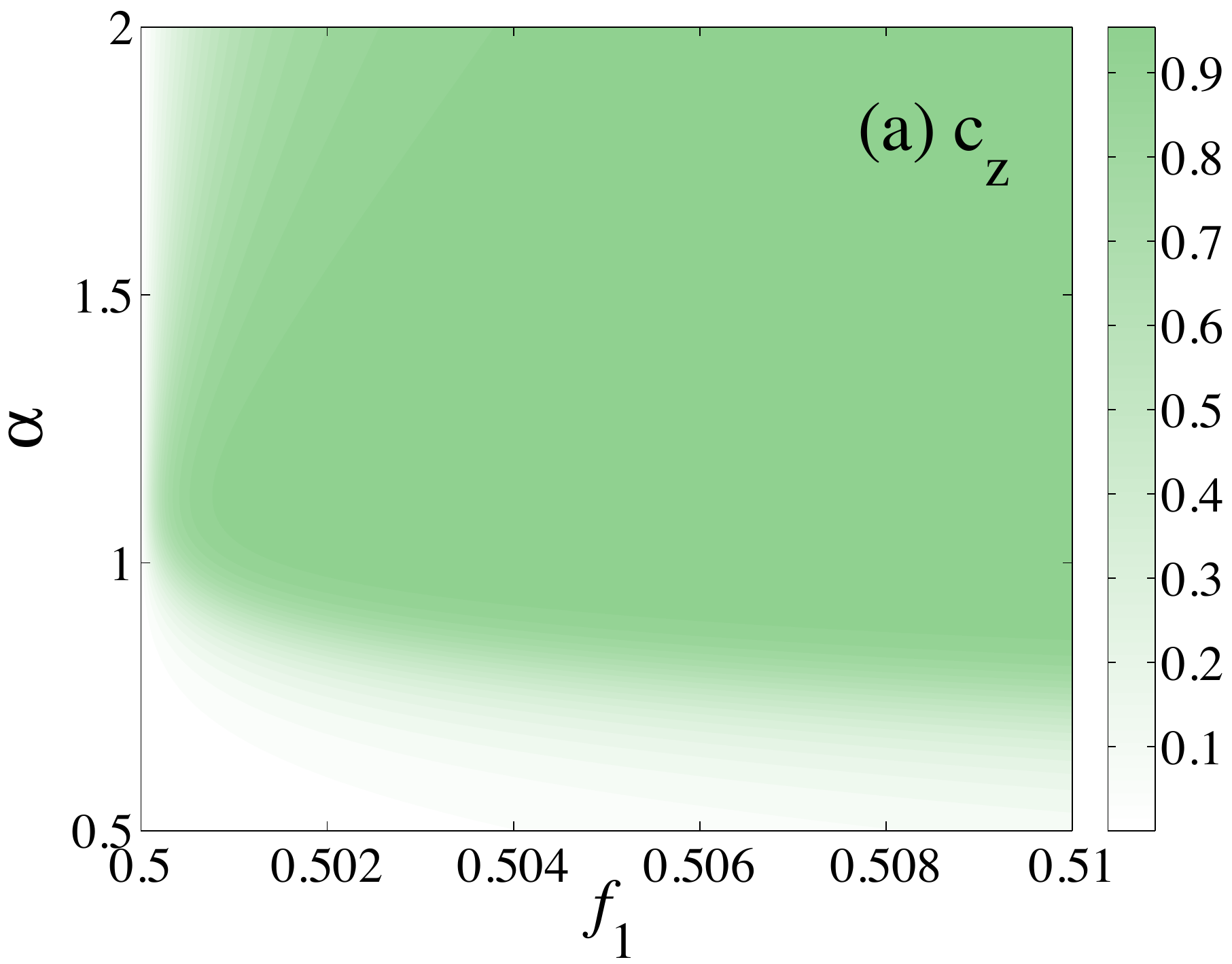}
\includegraphics[width=0.23\textwidth,height=0.2\textwidth]{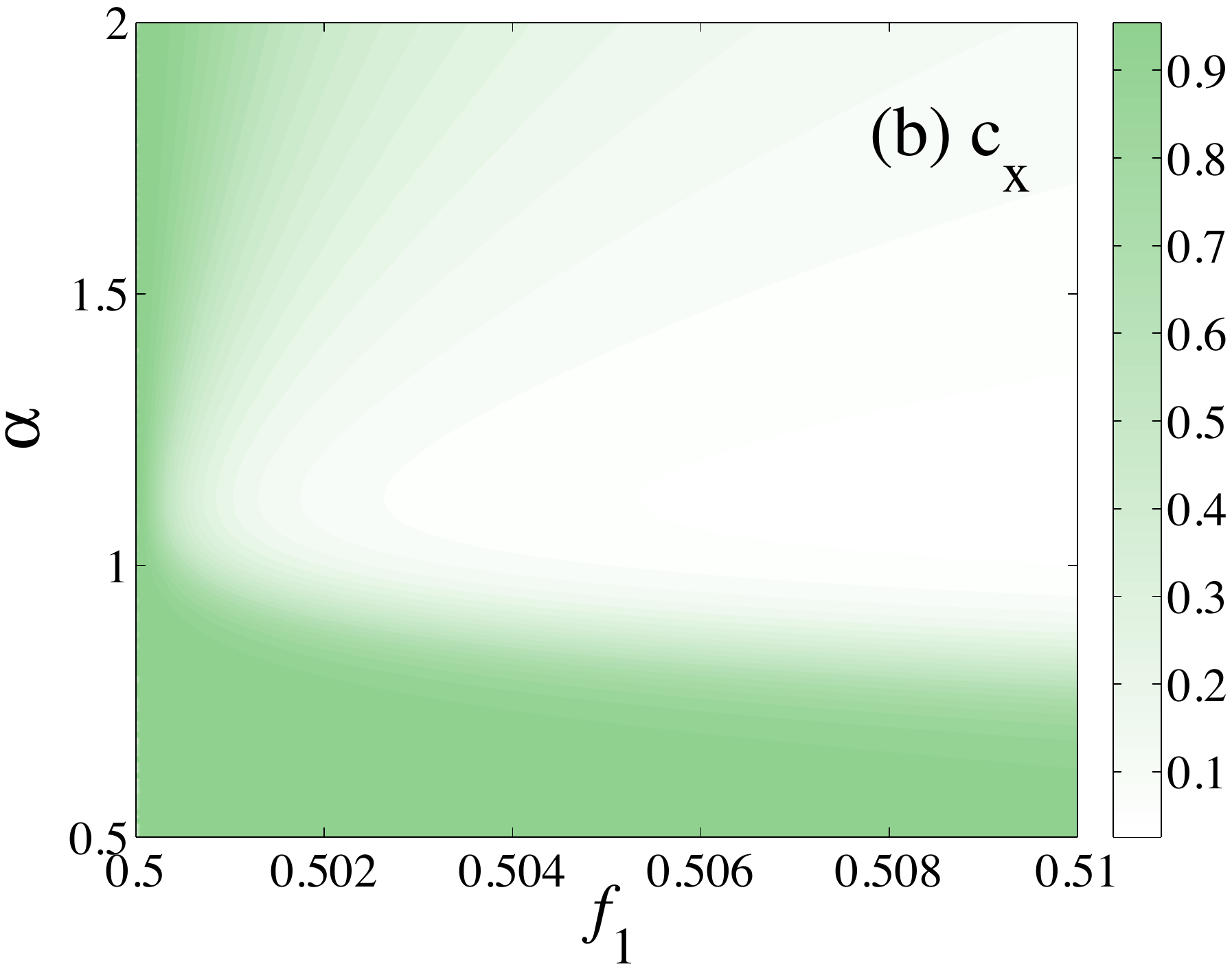}
\includegraphics[width=0.22\textwidth,height=0.2\textwidth]{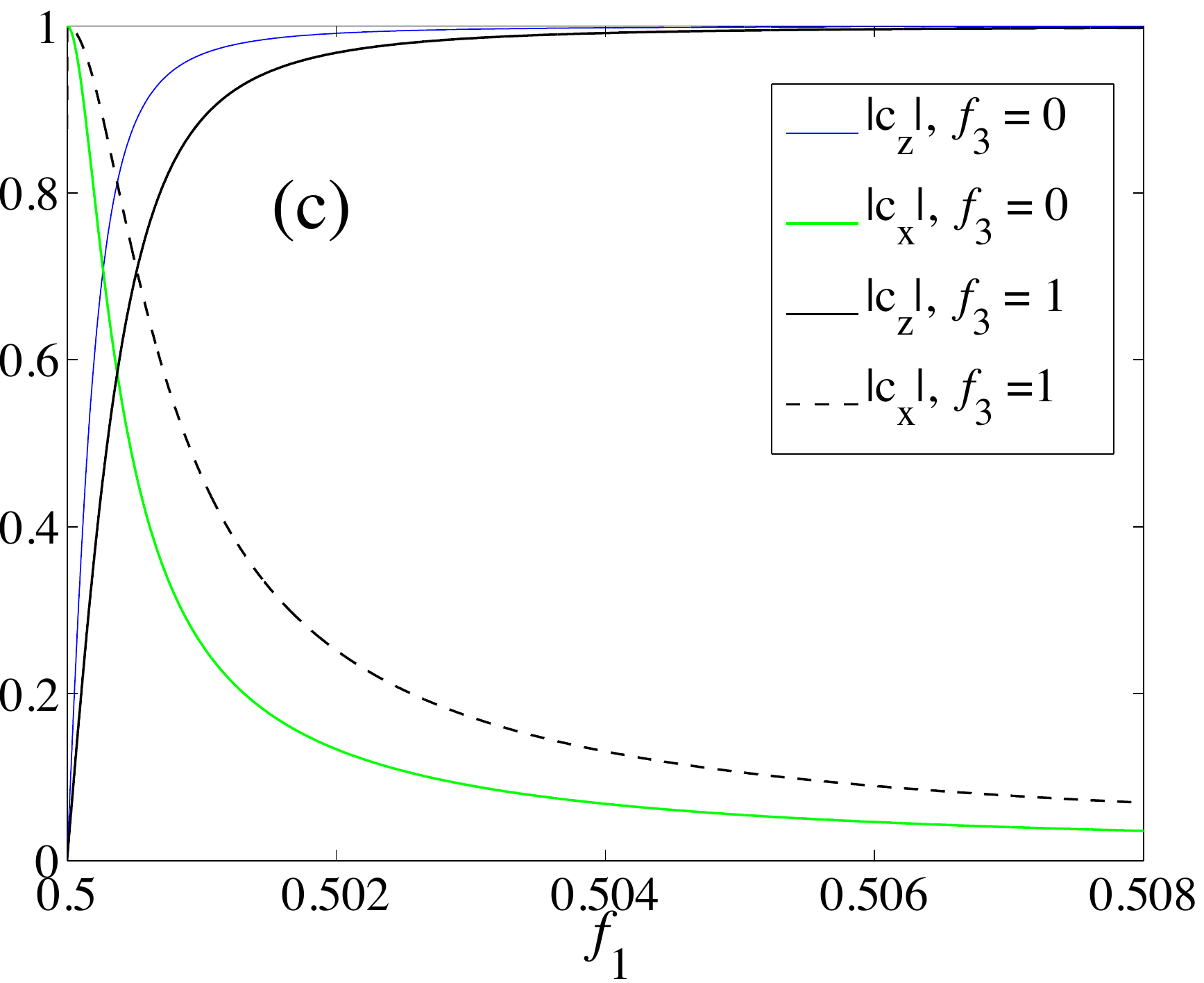}
\includegraphics[width=0.22\textwidth,height=0.2\textwidth]{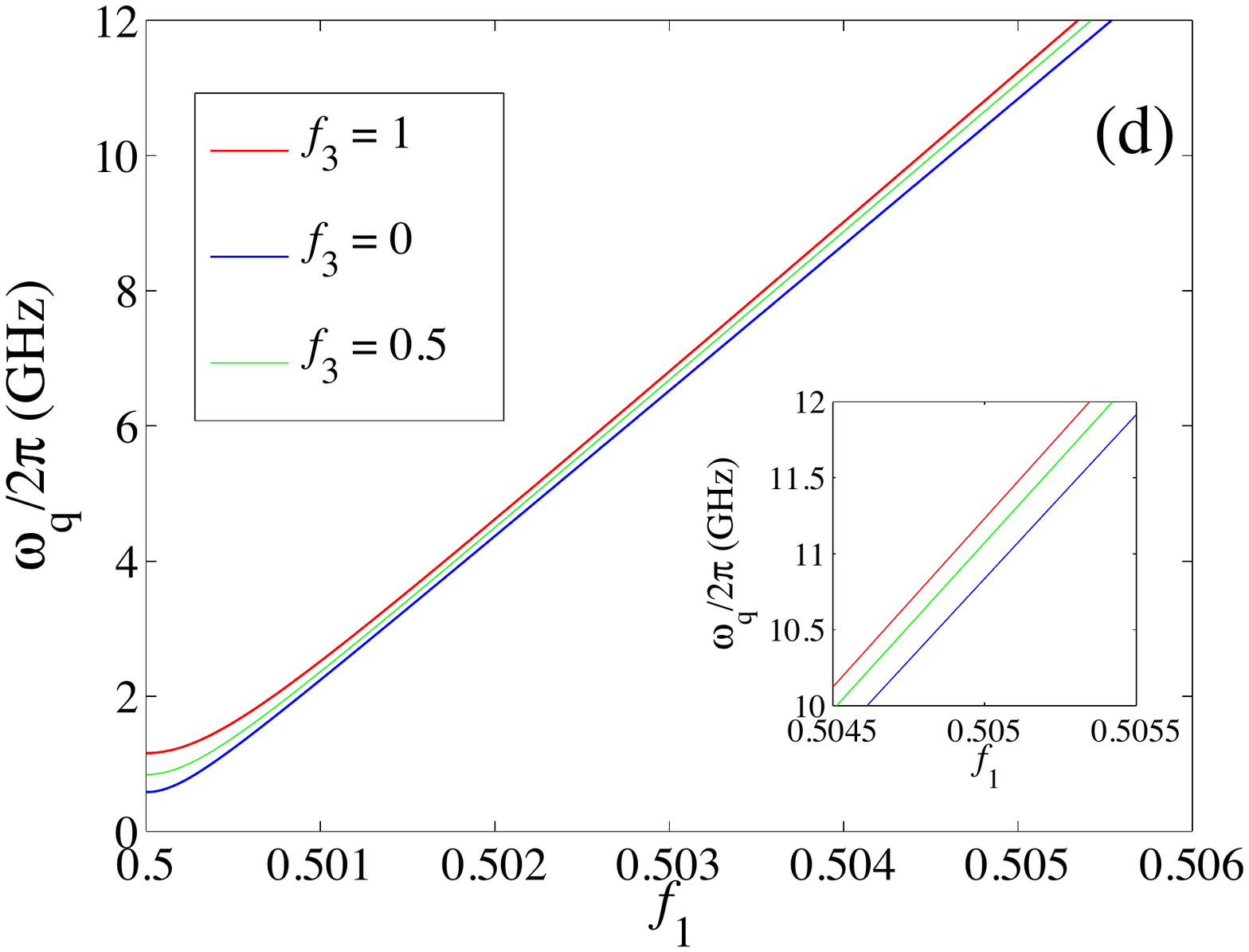}
\caption{\label{fig2}(Color online) (a,b) Couplings strength $c_z$ and $c_x$ as a function of the qubit junction size $\alpha = E_{J3}/E_{J1}$, and the external frustration parameter ${\it f}_1=\Phi_{1}/\Phi_0$, where $\Phi_0$ is the flux quantum, for a frustration $f_3=0$. (c) Coupling strengths for a qubit junction size $\alpha=1.2$, and for values $f_3=0$ and $f_3=1$. (d) Qubit spectrum for values $f_3=\{1,0,0.5\}$. In this simulation we have considered $E_J/h=221$~GHz, a junction size $\alpha_4=0.058$.}
\end{figure}

In Fig.~\ref{fig2}(a,b) we plot $c_{z}$ and $c_{x}$ as a function of the qubit junction size $\alpha$, and the frustration parameter $f_1$. They clearly show another characteristic of the setup, that is, the switching from transversal to longitudinal couplings depending on the external flux $\Phi_1$~\cite{Borja10}. In particular, when selecting a qubit junction size $\alpha=1.2$ and $f_1=0.505$, we obtain a large (small) contribution of longitudinal (transversal) coupling---see Fig.~\ref{fig2}(c) where $c_z$ and $c_x$ are depicted for a parameter $f_3=0$ and $f_3=1$.

From the diagonalization of qubit Hamiltonian for different values of the frustration parameter $f_3=\{0,1,0.5\}$, we estimate a qubit frequency $\omega_q/2\pi\sim \{10.94,11.25,10.99\}~{\rm GHz}$ [see Fig.~\ref{fig2}(d)]. Furthermore, considering a junction size $\alpha_4=0.058$, the qubit-resonator coupling strength is $g/\omega_r=\{0.446,-0.446,0\}$, reaching the USC regime. It is noteworthy to mention that our setup allows us to turn on/off the coupling $g$ as well as to change its sign, operations that may be carried out in times of the order of $0.1$~ns, or even less~\cite{Wilson,privatecomm}.

The ratio $g/\omega_r$ could change at will without affecting significantly the qubit properties required for the proposed protocol. For instance, for a junction size $\alpha_4=0.12$, a phase slip magnitude $\Delta\psi_1=0.0768$ and a resonator frequency $\omega_r/2\pi\sim8.01$~GHz (obtained from a central Josephson capacitance $C_{J6}\sim17$~fF), one obtains $g/\omega_r=0.509$. Also, with other choices of parameters, we can move from the USC to DSC regime and even the ratio of transversal to longitudinal coupling could be tuned. These examples, and particularly the model introduced by Eq.~(\ref{long_Hami2}), will be the basis to develop protocols for fast two-qubit gates.

{\it Fast two-qubit gate}.---The external tunability of the previous circuit is now exploited to propose an fast protocol for a two-qubit gate~\cite{JJ1}. In Ref.~\cite{Wang04} a scheme was studied, where a two-qubit CPHASE gate is produced by alternating between positive and negative values of the coupling strength $g$ for each qubit. In our architecture, this can be done simply by changing the flux $f_3$ from 0 to 1; however, this action will also increase the value of the undesired transversal coupling in Hamiltonian of Eq.~(\ref{long_Hamil}). For instance, in case of a junction size $\alpha_4=0.12$, we find $g/\omega_r=\{0.509,-0.509\}$ and $c_x=\{0.040,0.135\}$ for $f_3=\{0,1\}$. 

Hence, a more suitable protocol for the proposed architecture consists of: 
\begin{enumerate}
\item[Step 1.-]  The coupling $g_1$ is maximized (${f^{(1)}_3}=0$), whereas $g_2$ made exactly zero by tuning ${f^{(2)}_3}=0.5$. The system evolves for a period $\omega_r t_1\in(0,\pi/2]$.
\item[Step 2.-]  The coupling $g_2$ is maximized (${f^{(2)}_3}=0$), whereas $g_1$ made exactly zero by tuning ${f^{(1)}_3}=0.5$. The system evolves for a period $\omega_r t_2=\pi - \omega_r t_1$.
\item[Step 3.-]  Repeat Step 1.
\item[Step 4.-]  Repeat Step 2.
\end{enumerate}

We study first the ideal case, in which the transversal component of the coupling in Eq.~(\ref{long_Hamil}) is negligible, which could be achieved by tuning the fluxes $f_1=0.505$ (or larger) of each qubit. Then, the two-qubit Hamiltonian is
\begin{equation}
{\cal H} =\sum_i \frac{\hbar{\omega_q}_i}{2} \sigma^{(i)}_z +  \hbar \omega_r a^{\dag}a -   \sum_i  \hbar g_i (a + a^{\dag})    {\sigma^{(i)}_z }.
\label{long_Hamil_2}
\end{equation}
Under this Hamiltonian, the unitary evolution operator ${\cal U}_i$ corresponding to each step is
\begin{eqnarray}
{\cal U}_{1,3} &=& {\rm e}^{-i\frac{{\omega_q}_{1\uparrow}{\sigma^{(1)}_z}+{\omega_q}_{2\downarrow}{\sigma^{(2)}_z}}{2} t_1 }   {\rm e}^{-i\omega_r t_1 a^{\dag}a} {\cal D}\left(  \frac{ g_1}{\omega_r} \left(  {\rm e}^{i \omega_r t_1} -1 \right) {\sigma^{(1)}_z}  \right) \nonumber \\
{\cal U}_{2,4} &=& {\rm e}^{-i\frac{{\omega_q}_{1\downarrow}{\sigma^{(1)}_z}+{\omega_q}_{2\uparrow}{\sigma^{(2)}_z}}{2} t_2 }   {\rm e}^{-i\omega_r t_2 a^{\dag}a} {\cal D}\left(  \frac{ g_2}{\omega_r} \left( {\rm e}^{i \omega_r t_2} -1 \right) {\sigma^{(2)}_z}  \right) , \nonumber \\
\label{U}
\end{eqnarray}
where ${\cal D}\left( \beta \sigma_z \right)=\exp\left( \{\beta a^{\dag} - \beta^{*} a \} \sigma_z \right)$ is a controlled coherent displacement of the field and ${\omega_q}_{i\uparrow}$, ${\omega_q}_{i\downarrow}$ stands for the value of the qubit frequency when $g_i$ is maximum, zero, respectively. Finally, using ${\cal D}(\alpha){\cal D}(\beta)=e^{i \mathrm{Im}(\alpha\beta^*)}{\cal D}(\alpha+\beta)$ and $e^{-i\theta a^\dagger a}{\cal D}(\alpha)e^{i\theta a^\dagger a}={\cal D}(\alpha e^{-i\theta})$, the gate ${\cal U}=\prod_i{\cal U}_i$ is
\begin{eqnarray}
{\cal U} &=&  {\rm e}^{-i {(\omega_q}_{1\uparrow}t_1+{\omega_q}_{1\downarrow}t_2){\sigma^{(1)}_z} } \nonumber \\ & \times&  {\rm e}^{-i {(\omega_q}_{2\downarrow}t_1+{\omega_q}_{2\uparrow}t_2){\sigma^{(2)}_z}}  {\rm e}^{4i \sin{\omega_r t_1}\frac{g_1g_2}{\omega_r^2}  {\sigma^{(1)}_z}{\sigma^{(2)}_z}}\,.
\label{Y}
\end{eqnarray}

This gate ${\cal U}$ is equivalent to a CPHASE quantum gate~\cite{NC}, up to local unitary operations, provided $4 \sin{\omega_r t_1} g_1g_2/\omega_r^2 = \pi/4$ (notice that this condition requires $g_i/\omega_r$ to be in the ultrastrong regime). In the case of having a junction size $\alpha_4=0.12$, for which the coupling strength takes a value of $g_1/\omega_r=g_2/\omega_r=0.509$ and $\omega_r\sim2\pi\times8.01$ GHz, then $\omega_r t_1=0.86$ and the total gating time will be $t_{\rm gate}=2\pi/\omega_r=0.12$ ns. This is much shorter than typical coherence times in these systems, which are around $1~{\rm \mu s}$.

Deviations from perfect fidelity are expected if one accounts for undesired transverse coupling in Eq.~(\ref{long_Hamil}). For an initial state where both qubits are in state $\left| + \right\rangle = (\left| g \right\rangle + \left| e \right\rangle)/\sqrt2$ and the resonator in the vacuum, we can compute the fidelity of the state generated assuming that ${c^{(i)}_x}=0.040$ for each qubit at ${f^{(i)}_3}=0$. The fidelity of this state, with reduced density matrix $\rho$, as compared to the ideal $ \left | \psi \right\rangle  \left  \langle \psi  \right | $, for which transverse coupling is neglected, amounts to ${\cal F}=\left\langle \psi \right | \rho \left | \psi \right\rangle \geq 0.996$. This result is unchanged even if we include up to the third cavity mode in our ab initio calculation. For the sake of simplicity, we have considered instantaneous changes in the value of the fluxes $f_3$. In this sense, the scheme can be easily adapted to account for smooth time-dependent profiles in the value of the coupling strength of both qubits, provided the adequate interaction time and number of iterations, and that no overlap between the pulses occurs. Indeed, switching frequencies of about $10 - 80$ GHz are already available~\cite{Wilson,privatecomm}. This should allow the experimental realization of a high-fidelity fast CPHASE gate with subnanosecond operation time. 

{\it Discussion.}---Given the proposed CPHASE gate, based on the tunable qubit-resonator coupling in USC, we may consider the following extensions. (i) Multi-qubit entanglement and gate operations, such as realization of three-qubit Toffoli gates in the USC regime, faster than other schemes working in the strong coupling regime~\cite{MM,Wallraff_te,Wallraff_tt}. (ii) With the advantage of switchable coupling in both strength and orientation, we may think of generating Ising-type Hamiltonians for qubit arrays. (iii) By controlling the geometric-related flux values, we can increase the higher-order couplings and thus study the nonlinear dynamics in USC regime. (iv) The adjustable coupling also allows us to couple to slower measurement devices.

{\it Conclusion.}---We have proposed a realistic scheme for implementing an fast two-qubit CPHASE gate in current circuit-QED technology. The gate may work at subnanosecond time scales with fidelity ${\cal F}=0.996$. This proposal may lead to a significant improvement in the operating time with respect to standard circuit-QED scenarios, as well as microwave/optical cavity QED systems, together with the reduction of the large resource requirement for fault-tolerant quantum computing.

\begin{acknowledgments}
We thank P. Forn-D\'iaz, B. Peropadre, J. J. Garc\'ia-Ripoll, M. S. Kim, and C. M. Wilson, for useful discussions. This work was supported by the National Research Foundation and the Ministry of Education, Singapore, Juan de la Cierva Program, MICINN FIS2009-12773-C02-01, Basque Government IT472-10, SOLID, and CCQED European projects.
\end{acknowledgments}


\begin{thebibliography}{99}

\bibitem{NC} M. A. Nielsen and I. L. Chuang, {\it Quantum Computation and Quantum Information} (Cambridge University Press, Cambridge, 2000).

\bibitem{QC} T. D. Ladd, F. Jelezko, R. Laflamme, Y. Nakamura, C. Monroe, and J. L. O'Brien, Nature {\bf 464}, 45 (2010).

\bibitem{Blais04} A. Blais, R. S. Huang, A. Wallraff, S. M. Girvin, R. J. Schoelkopf, Phys. Rev. A 69, 062320 (2004).

\bibitem{Wallraff04} A. Wallraff {\it et al.}, Nature (London) 431, 162 (2004).

\bibitem{Chiorescu04} I. Chiorescu {\it et al.}, Nature (London) 431, 159 (2004).

\bibitem{IonsReview} D. Leibfried, R. Blatt, C. Monroe, and D. Wineland, Rev. Mod. Phys. {\bf 75}, 281 (2003).

\bibitem{Optical} P. Kok, W. J. Munro, K. Nemoto, T. C. Ralph, J. P. Dowling, and G. J. Milburn, Rev. Mod. Phys. {\bf 79}, 135 (2007).

\bibitem{Niemczyk10} T. Niemczyk, F. Deppe, H. Huebl, E. P. Menzel, F. Hocke, M. J. Schwarz,	 J. J. Garc\'ia-Ripoll, D. Zueco, T. H\"ummer, E. Solano, A. Marx, and R. Gross, Nature Phys. {\bf 6}, 772 (2010).

\bibitem{Pol10} P. Forn-D\'iaz, J. Lisenfeld, D. Marcos, J. J. Garc\'ia-Ripoll, E. Solano, C. J. P. M. Harmans, and J. E. Mooij, Phys. Rev. Lett. {\bf 105}, 237001 (2010).

\bibitem{Borja10} B. Peropadre, P. Forn-D\'iaz, E. Solano, and J. J. Garc\'ia-Ripoll, Phys. Rev. Lett. {\bf 105}, 023601 (2010).
    
\bibitem{Plantenberg07} J. H. Plantenberg {\it et al.}, Nature {\bf 447}, 836 (2007).

\bibitem{DiCarlo09} L. DiCarlo {\it et al.}, Nature {\bf 460}, 240 (2009).

\bibitem{Yamamoto10} T. Yamamoto {\it et al.}, Phys. Rev. B {\bf 82}, 184515 (2010).

\bibitem{Bialczak10} R. C. Bialczak {\it et al.}, Nature Phys. {\bf 6}, 409 (2010).

\bibitem{Geller10} M. R. Geller, E. J. Pritchett, A. Galiautdinov, and J. M. Martinis, Phys. Rev. A {\bf 81}, 012320 (2010).

\bibitem{Haack10} G. Haack, F. Helmer, M. Mariantoni, F. Marquardt, and E. Solano, Phys. Rev. B {\bf 82}, 024514 (2010).

\bibitem{3qbit}  L. DiCarlo {\it et al.}, Nature {\bf 467}, 574 (2010); M. Neeley {\it et al.}, Nature {\bf 467}, 570 (2010);

\bibitem{MM} M. Mariantoni {\it et al.}, Science {\bf 334}, 61 (2011).

\bibitem{Wallraff_te} V. M. Stojanovic, A. Fedorov, C. Bruder, and A. Wallraff. arXiv:1108.3442.

\bibitem{Ciuti05} C. Ciuti, G. Bastard, and I. Carusotto, Phys. Rev. B {\bf 72}, 115303 (2005).

\bibitem{Bourassa09} J. Bourassa, J. M. Gambetta, A. A. Abdumalikov, Jr., O. Astafiev, Y. Nakamura, and A. Blais, Phys. Rev. A {\bf 80}, 032109 (2009).

\bibitem{JJ1} See, as a reference, the proposed fast gates in trapped ions: J. J. Garc\'ia-Ripoll, P. Zoller, and J. I. Cirac, Phys. Rev. Lett. {\bf 91},157901 (2003); J. J. Garc\'ia-Ripoll, P. Zoller, and J. I. Cirac, Phys. Rev. A {\bf 71}, 062309 (2005).  

\bibitem{JJ2} W. C. Campbell et al., Phys. Rev. Lett. {\bf 105}, 090502 (2010); Chang-Yong Chen, Commun. Theor. Phys. {\bf 56}, 91 (2011). 

\bibitem{Braak11} D. Braak, Phys. Rev. Lett. {\bf 107}, 100401 (2011).

\bibitem{Solano11} E. Solano, Physics {\bf 4}, 68 (2011).

\bibitem{Wang04} Y. D. Wang, P. Zhang, D. L. Zhou, and C. P. Sun, Phys. Rev. B {\bf 70}, 224515 (2004).

\bibitem{Wang09} Y. D. Wang, A. Kemp, and K. Semba, Phys. Rev. B {\bf 79}, 024502 (2009).

\bibitem{Wang10} Y. D. Wang, S. Chesi, D. Loss, and C. Bruder, Phys. Rev. B {\bf 81}, 104524 (2010).

\bibitem{Ciuti11} P. Nataf and C. Ciuti, Phys. Rev. Lett. {\bf 107}, 190402 (2011).

\bibitem{Mazo99} T. P. Orlando, J. E. Mooij, L. Tian, C. H. Van der Wal, L. Levitov, S. Lloyd, and J. J. Mazo, Phys. Rev. B {\bf 60}, 15398 (1999).

\bibitem{Casanova2010} J. Casanova, G. Romero, I. Lizuain, J. J. Garc\'ia-Ripoll, and E. Solano, Phys. Rev. Lett. {\bf 105}, 263603 (2010).

\bibitem{Fault} E. Knill, Nature {\bf 434}, 39 (2005); R. Raussendorf, J. Harrington, Phys. Rev. Lett. {\bf 98}, 190504 (2007).

\bibitem{Abdumalikov08} A. A. Abdumalikov, O. Astafiev, Y. Nakamura, Y. A.Pashkin, and J. S. Tsai, Phys. Rev. B {\bf 78}, 180502 (2008).

\bibitem{ABC}
These values are $A = (\tilde{C}_r(1+\alpha+2 \alpha_4)+ 2 \alpha_4 \,\delta_1^2 \,C_J (1+\alpha))/(\tilde{C}_r(1+(\alpha+2 \alpha_4))+2 \alpha_4 \delta_1^2 C_J (1+2 \alpha))$, $B = (2 \alpha \alpha_4 C_J \delta_1^2 + \tilde{C}_r(\alpha+2 \alpha_4))/(\tilde{C}_r(1+2 (\alpha+2 \alpha_4)) + 2 \alpha_4 \delta_1^2 C_J (1+2 \alpha))$, and $C = (2 \tilde{C}_r \alpha_4 \delta_1)/(\tilde{C}_r(1+2 (\alpha+2 \alpha_4)) + 2 \alpha_4 \delta_1^2 C_J (1+2 \alpha))$.

\bibitem{Wilson} C. M. Wilson, G. Johansson, A. Pourkabirian, J. R. Johansson, T. Duty, F. Nori, P. Delsing, Nature {\bf 479}, 376 (2011);

\bibitem{privatecomm} C. M. Wilson, private communication.

\bibitem{Wallraff_tt} A. Fedorov, L. Steffen, M. Baur, and A. Wallraff 	arXiv:1108.3966.

\end{thebibliography}
\end{document}